# Writable spin wave nanochannels in an artificial-spin-ice-mediated ferromagnetic thin film

Jianhua Li[1,2,3], Wen-Bing Xu[1,2], Wen-Cheng Yue[1,2], Zixiong Yuan[1,2], Tan Gao[1,2], Ting-Ting Wang[1,2], Zhi-Li Xiao[4,5], Yang-Yang Lyu[2], Chong Li[1,2], Chenguang Wang[1,2], Fusheng Ma[6,*], Sining Dong[1,2], Ying Dong[7], Huabing Wang[2,8], Peiheng Wu[2,8], Wai-Kwong Kwok[4] and Yong-Lei Wang[1,2,8*]

[1] *School of Electronic Science and Engineering, Nanjing University, Nanjing, 210023, China*

[2] *Research Institute of Superconductor Electronics, Nanjing University, Nanjing, 210023, China*

[3] *School of Physics and Electronic Electrical Engineering, Huaiyin Normal University, Huaian, 223300, China*

[4] *Materials Science Division, Argonne National Laboratory, Argonne, IL 60439, USA*

[5] *Department of Physics, Northern Illinois University, DeKalb, IL 60115, USA*

[6] *School of Physics and Technology, Nanjing Normal University, Nanjing 210046, China*

[7] *Research Center for Quantum Sensing, Zhejiang Laboratory, Hangzhou, Zhejiang, 311100, China*

[8] *Purple Mountain Laboratories, Nanjing, China*

* Correspondence to: phymafs@njnu.edu.cn; yongleiwang@nju.edu.cn

## Abstract

**Magnonics, which employs spin-waves to transmit and process information, is a promising venue for low-power data processing. One of the major challenges is the local control of the spin-wave propagation path. Here, we introduce the concept of writable magnonics by taking advantage of the highly flexible reconfigurability and rewritability of artificial spin ice systems. Using micromagnetic simulations, we show that globally switchable spin-wave propagation and the locally writable spin-wave nanochannels can be realized in a ferromagnetic thin film underlying an artificial pinwheel spin ice. The rewritable magnonics enabled by reconfigurable spin wave nanochannels provides a unique setting to design programmable magnonic circuits and logic devices for ultra-low power applications.**

1. The challenge to Moore's law due to the physical limitations of electronic devices, has propelled magnonics and/or spintronics as promising venues to transmit and process information with high performance[1-10]. The current research in magnonics is to pave a promising way for post-Moore information and communications technology[1-10]. In magnonics or magnon spintronics, spin waves (or magnons) instead of electrical charge are utilized as carriers of information, circumventing energy waste from Joule heating induced by charge transport[1-10]. Compared to an electromagnetic wave at the same frequency range, the wavelengths of a spin wave (SW) are several orders of magnitude shorter, enabling magnonic devices at the nanometer scale[1,11]. In the past few years, various magnonic devices have been proposed, such as, spin-wave directional coupler[12-13], magnon transistor[14] and majority gate[15]. A major challenge in magnonic applications is to effectively control the transmission paths of a spin wave. Efficient channeling and steering of spin waves are crucial in the route towards programmable magnonic devices.

2. Artificial spin ices (ASIs) are magnetic metamaterials comprised of dipolar coupled magnetic nanoislands placed in a certain geometrical arrangement[16]. They are natural analogs of magnonic crystals for manipulating spin waves[4,10,17-31]. ASIs were initially introduced as a macroscopic model system mimicking the atomic spin frustration in the rare earth pyrochlore[16]. The field has since matured into an exciting research area[-21] for exploration of fascinating physical phenomena, such as geometrical frustration[16, 32], magnetic monopoles[33-34] and Coulomb phases[35]. Its reconfigurable aspect has allowed unique functionalities for applications, such as data storage[18,21] and advanced computations[36-37]. Recent studies demonstrate that artificial spin ices possess rich mode spectra and tunable band structures in the GHz frequency range[17,26-30], which can be tailored by tuning the geometry and the magnetization state of ASIs for desired functionalities. These results indicate that ASIs can have great potential for programmable magnonics.

**3**. Here, we design a writable magnonic device which consists of a chiral or pinwheel ASI on top of a soft ferromagnetic thin film [Fig. 1(a)]. Through micromagnetic simulations we demonstrate that nanocale channels of spin wave transmission can be produced in the thin film under zero bias magnetic field. These nanochannels are regulatable by tuning the magnetization configuration of the top pinwheel ASI. The reconfigurability and rewritability of the ASI[38-39] enables these nanochannels in the ferromagnetic film to be globally switchable and locally writable. Our conceptual demonstration of writable magnonics could pave a flexible way for designing programmable spin-wave-based logic devices and circuits.

**4.** Recently, ferromagnetic thin film was introduced as an underlayer for an ASI system to tune the latter's frustration[40] and spin wave modes[23]. Here, we demonstrate the possibility for realizing writable magnonics in an ASI-mediated magnetic film [Fig.1(a)], by taking advantage of the reconfigurability and rewritability of the ASI to create and control nano-scale spin wave paths in the underlayer. For our simulations, we chose the recently developed pinwheel ASI pattern, conceived by rotating each nanobar magnet in a square lattice by 45º around its center[41-45] as illustrated in Fig.1(b). The pinwheel ASI manifests interesting properties, such as emergent charility[41] and domain wall topology[43]. One of its unique features is the formation of reconfigurable parallel chains of magnetic charges[43,45], which has been used to design programmable superconducting electronic devices[45]. In our simulations, we use elliptical nanobar magnets with dimensions of length

$L$ = 300 nm, width $W$ = 80 nm and thickness $T$ = 20 nm [Fig.1(b)]. At these dimensions, the nanobar magnets can maintain a single-domain state with a bistable remnant magnetization pointing along its long axis at room temperature, and can be locally manipulated using a magnetic writing technique[38]. The underlying ferromagnetic film [Fig.1(a)] is 10 nm thick. The micromagnetic simulations were carried out with Mumax3[46]. Details of the micromagnetic simulations are described in the supplementary material. We use the magnetic parameters of Permalloy ($Ni_{80}Fe_{20}$) for both the top nanobar magnets and the underlaying ferromagnetic film. Permalloy is a metallic soft ferromagnetic material with large saturation magnetization, low damping constant, and negligible magnetocrystalline anisotropy. The used parameters are as follows[46-48]: saturation magnetization $M_S$ = 8.6 × $10^5$ A/m; damping $\alpha$ = 0.01; and exchange constant $A_{ex}$ = 13 × $10^{-12}$ J/m. The magnetocrystalline anisotropy is not considered.

**5.** The pinwheel ASI has four-fold degenerate ferromagnetic orders, which are easily tunable by applying an in-plane external magnetic field[43,45]. The black arrows in Figs. 2(a) and 2(b) show two of its degenerate magnetic configurations. Since an ASI made of 20 nm thick nanomagnets is athermal, these ordered configurations are stable under zero bias magnetic field at room temperature. The corresponding magnetization distribution of the underlying ferromagnetic film for these two ordered ASI configurations, displays vertical [Fig. 2(a)] and horizontal [Fig. 2(b)] meandering stripes of domain and domain walls in the absence of an external field, respectively. The two states are obtained by polarizing the sample with horizontal (for vertical domain state) and vertical (for horizontal domain state) in-plane magnetic fields, respectively. Therefore, it is easy to switch between the two states by tuning the magnetic configuration of the ASI pinwheel pattern. Furthermore, due to the athermal nature of the ASI, the in-plane magnetic field can be removed once the nanomagnets are polarized, so that the device can work at zero bias magnetic field.

6. Previous investigations have shown that domain walls can serve as excellent magnonic waveguide channels for spin wave propagation[49]. This suggests that the reconfigurable domains and/or domain walls in the underlying magnetic film mediated by the pinwheel ASI could be considered as a perfect magnonic crystal for in-situ control of spin wave transmissions. To examine the spin wave dynamics in our continuous magnetic film, we calculate the mode spectra for the vertical and horizontal domain states, as shown in Fig. 2(c). The excitation of the external magnetic field pulse is along $+x$ (horizontal) direction in the sample plane. The spectra are calculated for both the top pinwheel ASI pattern and the underlayer film with periodic boundary conditions in the $x$-$y$ plane. Both spectra show prominent peaks ($V_1$ for vertical domain state and $H_1$ for horizontal domain state) of eigenmodes with amplitudes much larger than the other minor modes $V_2$, $H_2$ and $H_3$ [Fig. 2(c)]. However, the spectra of the vertical and horizontal domain states are quite distinct, i.e., in both their mode frequencies and amplitudes. The prominent eigenmodes of the horizontal/vertical domain states are at 3.71/4.30 GHz while the amplitude of the horizontal horizontal domain state is nearly three times higher than that of the vertical domain state.

7. To investigate the origin and difference of the spectrum eigenmodes between the two domain states, we calculate the spatial mode profiles of spin waves with various frequencies in the horizontal/vertical domain states. The spectrum maps of the top ASI and the underlayer film (see Figs. S1 and S2 in the supplementary material) clearly show that these eigenmodes are dominated by the responses from the underlayer film. Therefore, in the following we will mainly present the spin wave responses from the underlayer film. The frequency evolutions of the spatial mode profiles for the two domain states is shown in Videos 1 and 2, respectively. In Figs. 2(d) and 2(e), we show the spatial mode profiles of spin wave at $H_1$ = 3.71 GHz in the underlayer film for the two domain states, respectively. The spatially resolved maps for the other modes can be found in Figs. S1 and S2 in the supplementary material. The vertical domain state displays discrete, weak and diffusive modes [Fig. 2(d)]. In contrast, the horizontal domain state shows continuous, strong and clear nanochannels of the mode at the domain walls. This agrees well with the domain wall based magnonic waveguides[49]. The reason for the absence of strong mode channels at the domain walls for the vertical domain state is that the magnetic moments in the domain walls are all along the *x* (horizontal) direction [Fig. 2(a)], which is the same with the excitation orientation of the microwave pulse. On the other hand, the magnetic moments of the domain walls in the horizontal domain state are all in the vertical direction [Fig. 2(b)], which is perpendicular to the microwave pulse direction. Therefore, strong resonance of moment precession in the domain walls is produced. We also simulated the spin wave excitations along *y* and *z* directions (see Figs. S3 and S4 in the supplementary material). The spectra of the horizontal/vertical states with *y*-direction excitation are reversed as compare to those with *x*-direction excitation. The spectra of the two states with *z* direction excitation are exactly the same. The detailed analysis for the spectra with excitations along *y* and *z* directions can be found in the supplementary material. The significantly different responses between vertical and horizontal domain states are indicative of a perfect reconfigurable magnonic crystal, in which the spin wave transmission can be conveniently switched on and/or off by tuning the magnetic configurations of the top ASIs. The spectra excited by in-plane microwave can be conveniently realized by patterning samples on top of a coplanar waveguide (CPW)[24-27,30] or mounting samples on a CPW chip with a flip-chip technique in broadband ferromagnetic resonance (FMR) experiments[29]. Thus, our proposed reconfigurable magnonic crystal with tunable magnonic spectra could be directly realized by FMR measurements.

8. To further demonstrate the reconfigurable propagation of spin waves, we simulate a larger area of the sample. We apply a continuous excitation field (a sinusoidal field at $H_1=$ 3.71GHz along the $x$ direction) in a 100 nm wide center line, as shown in Figs. 3(a) and 3(c). The detailed simulation protocol is described in the supplementary material. Spin wave propagations in the two magnetic states can be directly revealed from the temporal magnetic moment mappings as in Videos 3 and 4, respectively. We extract the spatial maps of the spin wave amplitude in the underlayer film, as displayed in Figs. 3(b) and 3(d). It shows that the transmission of spin wave is limited to a narrow range near the center excitation line and cannot propagate over a long distance in the horizontal direction [Video 3 and Fig. 3(b)]. This is consistent with the weak spin wave mode in the vertical domain state under an excitation in the $x$ direction [Fig. 2(d)]. In contrast, in the horizontal domain state with strong mode channels [Fig. 2(e)], the spin wave transmits over a much longer distance along the horizontal nanochannels [Video 4 and Fig. 3(d)]. We also find that spin wave propagates asymmetrically on the two sides of the center excitation line along a single nanochannel. This asymmetric spin wave propagation is most probably due to the symmetry broken of the magnetic moments distribution, e.g., the moments' distribution is not mirror symmetrical on the two sides of the center excitation line. Future studies are still needed to understand the microscopic origin of this effect. Our results directly demonstrate in-situ switchable spin wave propagation by reconfiguring the magnetization states of the top layer pinwheel ASI structure.

**9**. Recent state-of-the-art nanomagnetic writing techniques allow local control of magnetic configurations of the ASI using the tip of a magnetic force microscope[38-39]. This approach can be used to realize writable spin wave nanochannels in our ASI mediated magnetic film. To demonstrate this concept, we simulate the spin wave propagation in a composite magnetization state, in which the left/right sides of the excitation line is in the vertical/horizontal domain states, respectively [Fig. 4(a)]. The magnetic configurations of the composite state are shown in Fig. S5(a) in the supplemental material. Such a state is nearly impossible to achieve with a global magnetization process, but can be conveniently realized by the above mentioned magnetic writing techniques[38-39]. The spatially resolved mode of $H_1$ = 3.71 GHz displays the expected feature, i.e., the mode nanochannels appear only on the right side [Fig. 4(c)], resulting in a one-way spin wave propagation, as clearly displayed by Video 5 and Fig. 4(e). The polarity of such a one-way spin wave propagation can be reversed by writing a vertical domain state in the right side and a horizontal domain state in the left side [Video 6 and Fig. S6]. Furthermore, we can also design a magnetization state in which a narrow strip of horizontal domain state with three domain lines at the right is embedded within a vertical domain background [Fig. 4(b)]. In this case, the nanochannels of spectrum mode are limited to the narrow strip region [Fig. 4(d)], leading to spin waves propagating primarily to +$x$ direction along the horizontal strip [Video 7 and Fig. 4(f)]. These examples evidently demonstrate writable magnonics in our ASI mediated ferromagnetic film, by fully taking advantage of the global reconfiguration and local rewrite characteristic of the ASI system[38-39].

**10.** In summary, we proposed a reconfigurable magnonic crystal comprised of a pinwheel ASI pattern imprinted onto a soft ferromagnetic underlayer film. Using micromagnetic simulations, we numerically demonstrated that the vertical and/or horizontal meandering stripes of magnetic domains and domain walls in the underlayer film at zero bias magnetic field can be robustly introduced by writing the magnetic configuration of the top pinwheel ASI structure. Furthermore, the domain walls can sustain spin waves with well-defined frequency, providing nanochanneled waveguides that are switchable and writable. The working frequencies can be easily tailored by regulating the dimensions of ASI (see Figs. S7 and S8 in the supplementary material). Our results demonstrate a convenient and flexible approach to effectively guide and manipulate spin waves, highlighting a potential application of artificial spin ices for writable magnonics. This highly reconfigurable magnonic crystal would stimulate future magnonic applications, such as programmable spin wave circuits and logic devices for energy-efficient information and data processing.

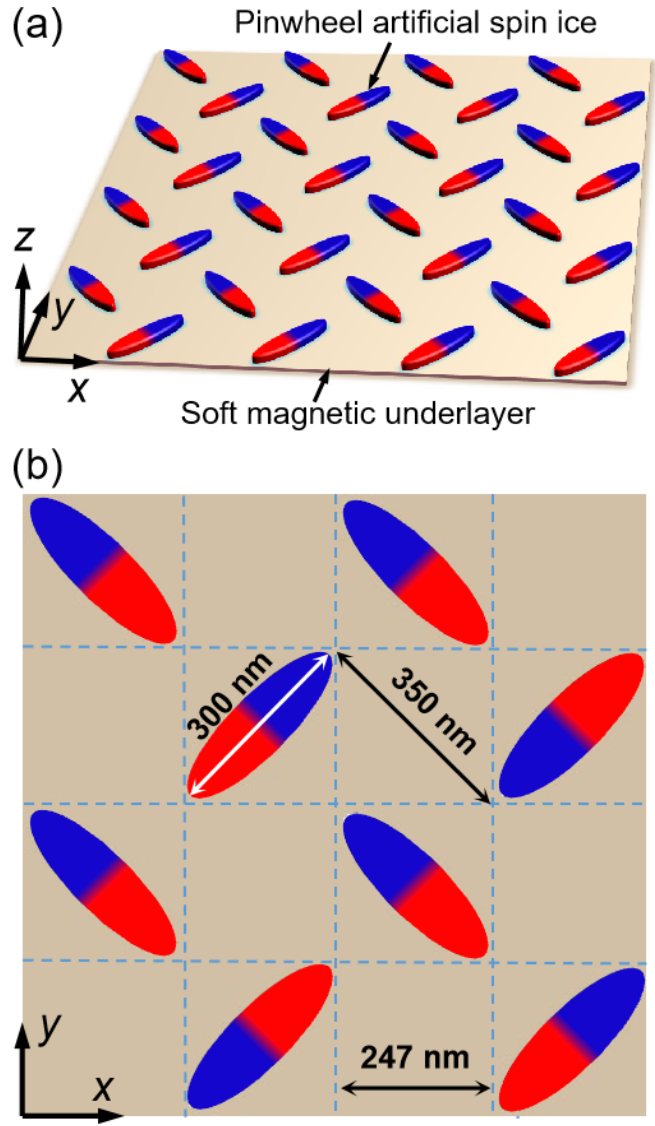

**Fig. 1 | Design of a system with writable spin wave nanochannels.** (a) Schematic showing a pinwheel artificial spin ice on top of an underlayer film. (b) Parameters of the top pinwheel artificial spin ice.

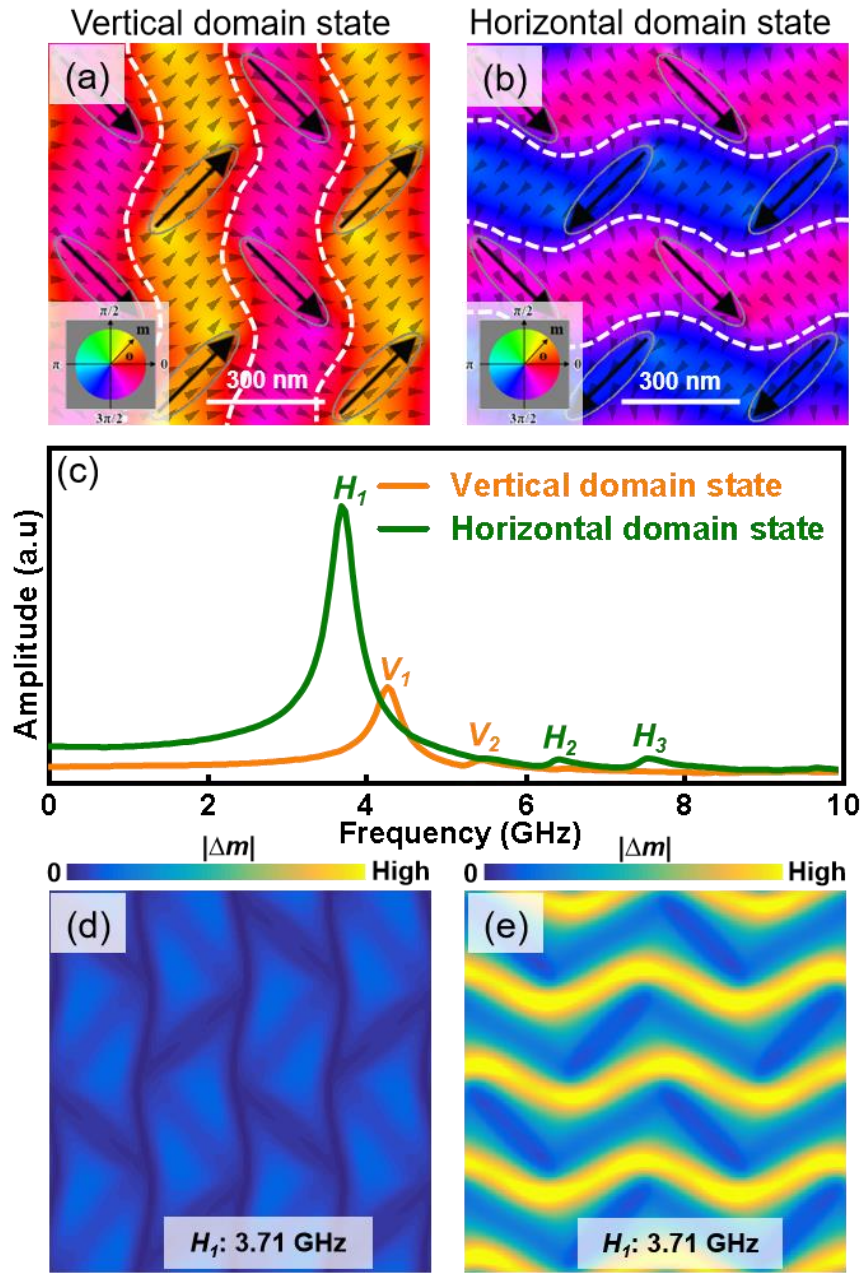

**Fig. 2 | Tunable magnetic structures and spin wave spectra. (a)** and **(b)** Vertical and horizontal stripe domains and domain walls in the soft magnetic underlayer film at zero external magnetic field, respectively. The black arrows show the corresponding magnetic configurations of the top pinwheel ASI structure. The white dash lines are domain walls. **(c)** The spectra for vertical and horizontal domain states excited by a field pulse along the x (horizontal) direction [defined in Fig. 1(b)]. The mode frequencies are $H_1$ = 3.71 GHz, $H_2$ = 6.50 GHz, $H_3$ = 7.62 GHz, $V_1$ = 4.30 GHz, $V_2$ = 5.47 GHz. **(d)** and **(e)** Spatially resolved modes at $H_1$ = 3.71 GHz for the vertical and horizontal domain states, respectively.

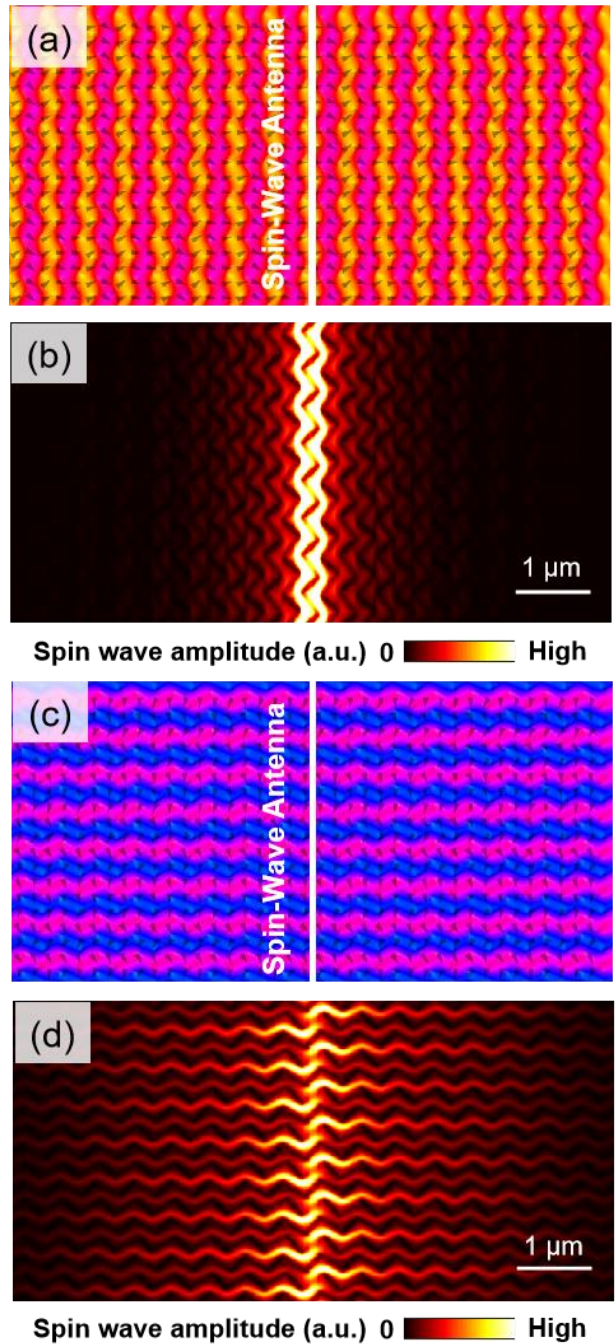

**Fig. 3 | Switchable spin wave propagation. (a)** and **(b)** Micromagnetic configuration and spin wave amplitude map of the vertical domain state, respectively, with the spin wave excitation in the white center line. **(c)** and **(d)** Micromagnetic configuration and spin wave amplitude map of the horizontal domain state, respectively.

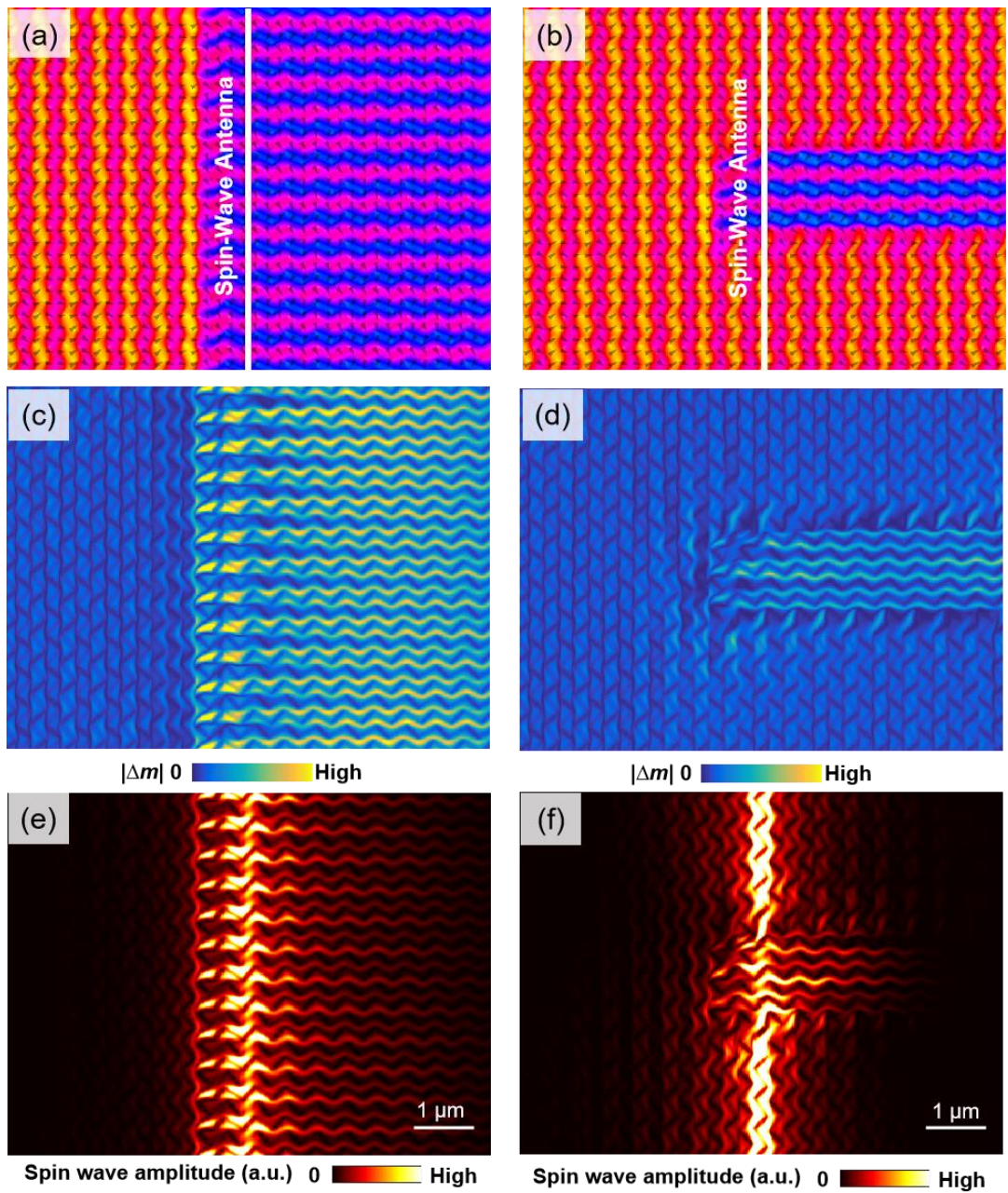

**Fig. 4 | Writable spin wave nanochannels. (a)** and **(b)** Composite magnetization state with both vertical and horizontal domain states, which can be realized by tuning the magnetization configurations of the ASI via currently available magnetic writing techniques (see main text). **(c)** and **(d)** Spatially resolved modes at $H_1$ = 3.71 GHz corresponding to (a) and (b). **(e)** and **(f)** Amplitude maps of propagating spin waves, corresponding to the states in (a) and (b), respectively. The spin waves are excited from the white center line shown in (a) and (b).

# Supplementary material

# Writable spin wave nanochannels in an artificial-spin-ice-mediated ferromagnetic thin film

Jianhua Li[1,2,3], Wen-Bing Xu[1,2], Wen-Cheng Yue[1,2], Zixiong Yuan[1,2], Tan Gao[1,2], Ting-Ting Wang[1,2], Zhi-Li Xiao[4,5], Yang-Yang Lyu[2], Chong Li[1,2], Chenguang Wang[1,2], Fusheng Ma[6,*], Sining Dong[1,2], Ying Dong[7], Huabing Wang[2,8], Peiheng Wu[2,8], Wai-Kwong Kwok[4] and Yong-Lei Wang[1,2,8*]

[1] *School of Electronic Science and Engineering, Nanjing University, Nanjing, 210023, China*

[2] *Research Institute of Superconductor Electronics, Nanjing University, Nanjing, 210023, China*

[3] *School of Physics and Electronic Electrical Engineering, Huaiyin Normal University, Huaian, 223300, China*

[4] *Materials Science Division, Argonne National Laboratory, Argonne, IL 60439, USA*

[5] *Department of Physics, Northern Illinois University, DeKalb, IL 60115, USA*

[6] *School of Physics and Technology, Nanjing Normal University, Nanjing 210046, China*

[7] *Research Center for Quantum Sensing, Zhejiang Laboratory, Hangzhou, Zhejiang, 311100, China*

[8] *Purple Mountain Laboratories, Nanjing, China*

\* Correspondence to: phymafs@njnu.edu.cn; yongleiwang@nju.edu.cn

Our micromagnetic simulations were carried out with Mumax3[46].

## 1. Simulations of the magnetization states and the spin wave spectra

The simulations are initialized with prescribed magnetization states for the pinwheel patterned ASI nanomagnets and with a uniform state for the underlayer ferromagnetic film. To eliminate relaxation ringing effects at low damping, the entire system relaxes to the ground state at zero magnetic field using a combination of high damping $\alpha = 1$ for 10 ns and a realistic damping of $\alpha = 0.01$ for 20 ns[34]. The simulations in Fig. 2 are conducted under a periodic boundary condition in the *x-y* plane. The cell size is 1.93 nm × 1.93 nm

× 5 nm. The system is excited by applying a uniform magnetic field pulse $B_{ext}$ = 2 mT for 50 ps along various directions described in the main text. We record the temporal magnetization $m(x, y, z, t)$ of all cells every 10 ps over the subsequent 20 ns. The fluctuation of each cell $m(x, y, z, t)$ is individually calculated via $\Delta m(x,y,z,t) = m(x,y,z,t) - m(x,y,z,t_0)$, where $m(x, y, z, t_0)$ corresponds to the ground magnetization state. The spectra are obtained from Fourier transformation of spatially averaged $\Delta m(x,y,z,t)$. The spin wave mode profile for a specific frequency can be mapped out as a function of position, $|\Delta m| = \sqrt{(\Delta m_x)^2 + (\Delta m_y)^2 + (\Delta m_z)^2}$.

## 2. Simulations of spin wave transmission

A larger sample of 10.15 μm × 10.15 μm is used for simulating the propagation of spin waves with open boundary conditions. The used cell size is 10 nm × 10 nm × 10 nm. In order to avoid boundary effect, we collect the data at the center region of the sample. A sinusoidal field $B_x = A\sin(2\pi ft)$ with frequency $f$ = 3.71 GHz and amplitude $A$ = 2 mT is locally applied to the 100 nm wide center region along the $x$ direction. We record the magnetization $m(x, y, z, t)$ every 10 ps. The fluctuations $m(x, y, z, t)$ were calculated for all cells via $\Delta m(x,y,z,t) = m(x,y,z,t) - m(x,y,z,t_0)$, where $m(x, y, z, t_0)$ corresponds to the ground state. The spin wave spatial amplitude map is calculated via

$$|A(t)| = \sqrt{(\Delta m(x,t)_{max})^2 + (\Delta m(y,t)_{max})^2 + (\Delta m(z,t)_{max})^2} \quad (1)$$

For the spin wave propagations shown in videos 1-5, the fluctuations $m(x, t)$ are calculated for all cells via $\Delta m(x,t) = m(x,t) - m(x,t_0)$, where $m(x, t_0)$ corresponds to the ground state. The spatial propagation of the spin wave is mapped out as a function of both $\Delta m(x,t)$ and position.

## 3. Spin wave spectra with *y*- and *z*-direction excitations

The spin wave excitations are not only determined by the magnetic configurations of the sample system, but also significantly influenced by the direction of the driving microwaves. We also simulate the spin wave spectra driven by microwave pulses along *y* and *z* directions. The results can be found in Figs. S3 and S4, respectively. The spectrum responses of both frequency [Figs. S3(a)] and spatial dependence [Figs. S3(b) and S3(c)] for the two domain states with *y*-direction excitation are reversed from the *x*-direction microwave pulse [Figs. 2(c)-2(e)]. In this case, the vertical domain state shows clear nanochannels of magnonic mode, which is not observed in the horizontal domain state. The spectra reversal in *y*-direction excitation can be easily understood by analyzing the orientation of the moments in domain walls and that of the driving microwave pulse, the same analysis used for *x*-direction excitations. For the case of out-of-plane excitation along the *z*-direction, which is perpendicular to the domain wall moments (in-plane) for both states, the spectra are exactly identical for both vertical and horizontal domain states, as shown in Fig. S4(a). The prominent modes for both states display continuous nanochannels (but with orthogonal directions for both states) under *z*-direction excitation [Fig. S4(b) and S4(c)]. However, the mode is much weaker than those from the in-plane excitations.

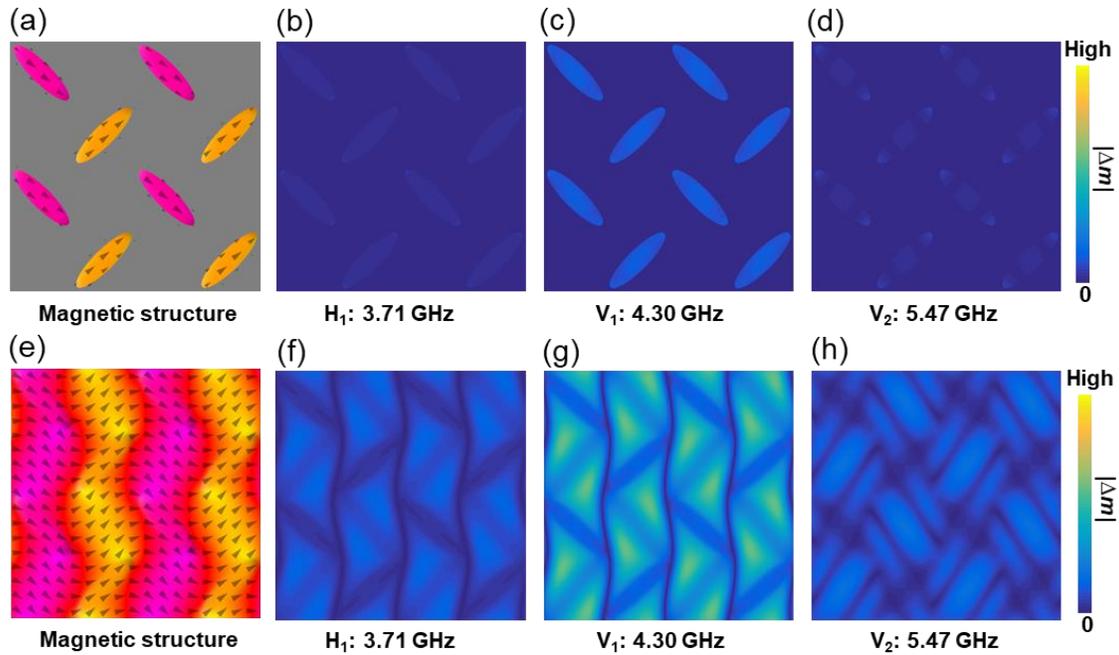

**Fig. S1 | Spin wave mode profiles for vertical domain state.** (a) Micromagnetic structure of the top pinwheel ASI in the ground state. (b)-(d) Spin wave mode profiles of different frequencies $H_1$, $V_1$, $V_2$ [as defined in Fig. 2(c)] for the top pinwheel ASI, respectively. **(e)** Micromagnetic structure of the underlayer film in the ground state. (f)-(h) Spin wave mode profiles of different frequencies $H_1$, $V_1$, $V_2$ in the underlayer film. The color scales of the mode maps are identical. The mode amplitudes of the underlayer are much stronger than those of the top ASI structure, suggesting that these eigenmodes are dominated by the responses from the underlayer film.

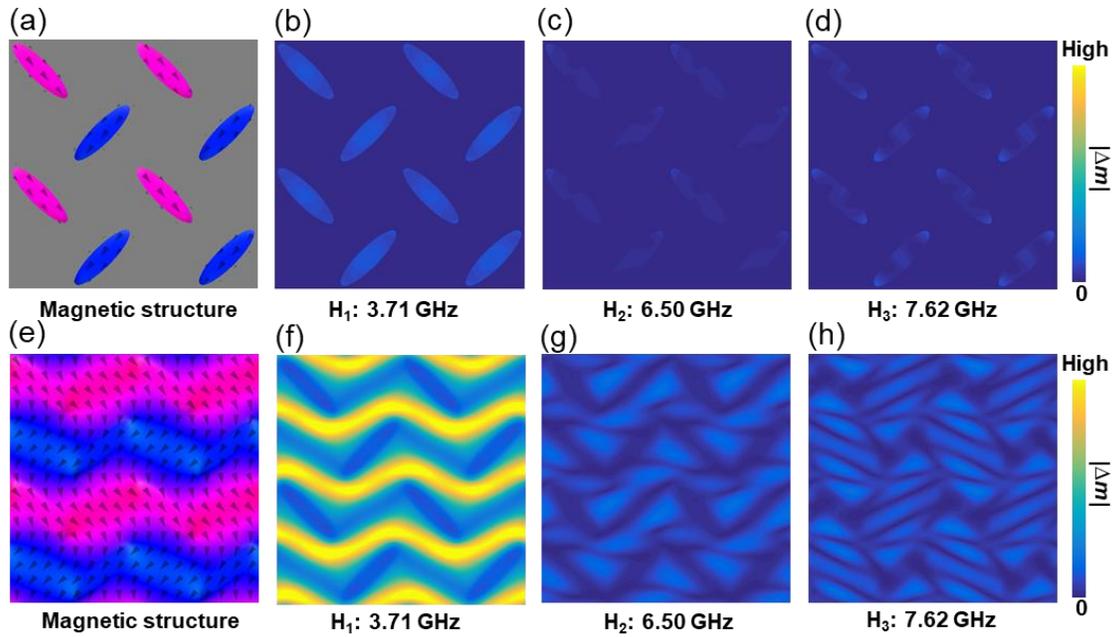

**Fig. S2 | Spin wave mode profiles for horizontal domain state. (a)** Micromagnetic structure of the top pinwheel ASI in the ground state. (b)-(d) Spin wave mode profiles of different frequencies $H_1$-$H_3$ [defined in Fig. 2(c)] for the top pinwheel ASI. **(e)** Micromagnetic structure of the underlayer film in the ground state. (f)-(h) Spin wave mode profiles of different frequencies $H_1$-$H_3$ in the underlayer film. The color scales of the mode maps are identical. The mode amplitudes of the underlayer film are much stronger than those of the top ASI, suggesting that these eigenmodes are dominated by the responses from the underlayer film.

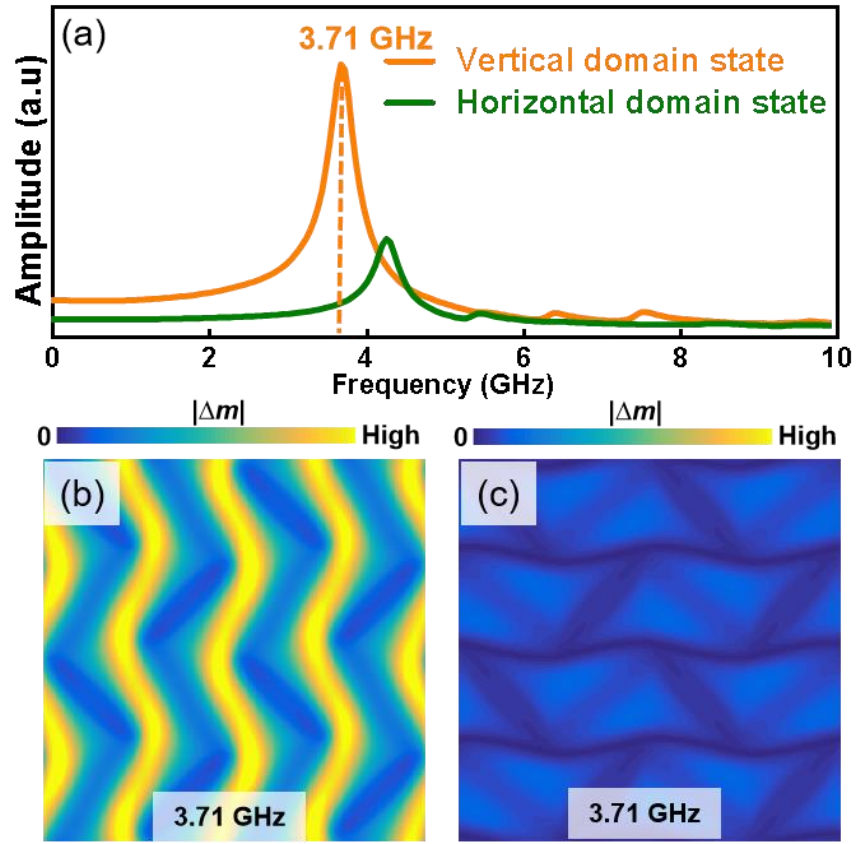

**Fig. S3 | Spin wave spectra with *y*-direction excitation. (a)** Spectra for vertical and horizontal domain states excited by a field pulse along *y* direction (defined in Fig. 1a). The axis scale is the same as that in Fig. 2(c). **(b)** and **(c)** Spin wave mode profile of frequency at 3.71 GHz corresponding to the two spectra in (a) and for the vertical and horizontal domain states, respectively. The color scale is the same as that in Figs. 2(d) and 2 (e).

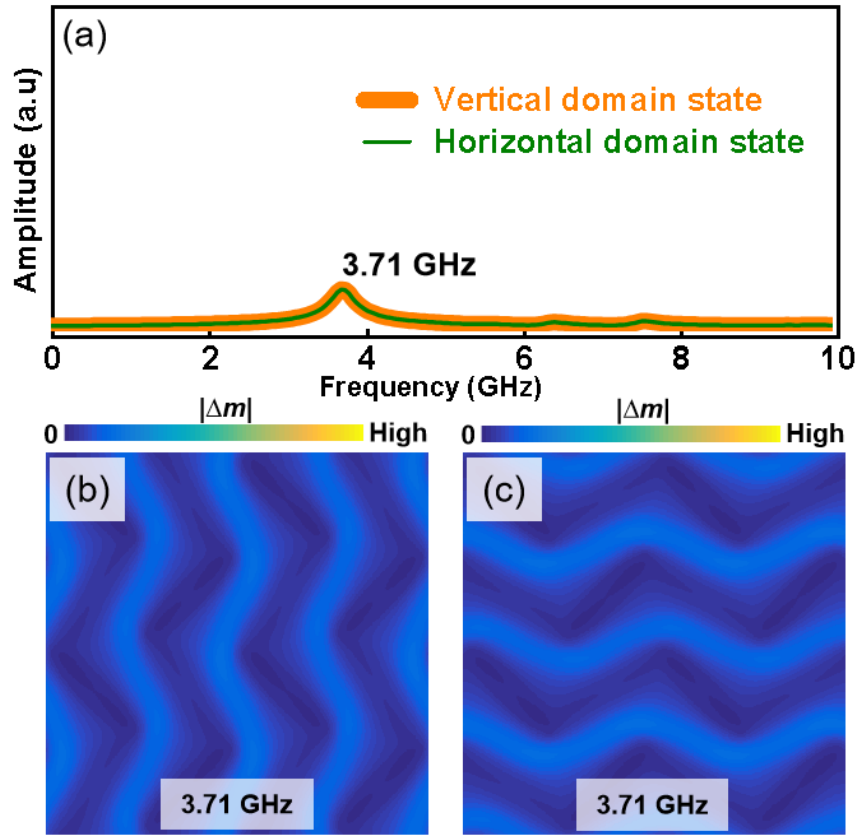

**Fig. S4 | Spin wave spectra with *z*-direction excitation. (a)** The spectra for vertical and horizontal domain states excited by a field pulse along the *z* direction (defined in Fig. 1a). The axis scale is the same as that in Fig. 2(c). **(b)** and **(c)** Spin wave mode profile of frequency 3.71 GHz corresponding to the two spectra in (a) and for the vertical and horizontal domain states, respectively. The color scale is the same as that in Figs. 2(d) and 2 (e).

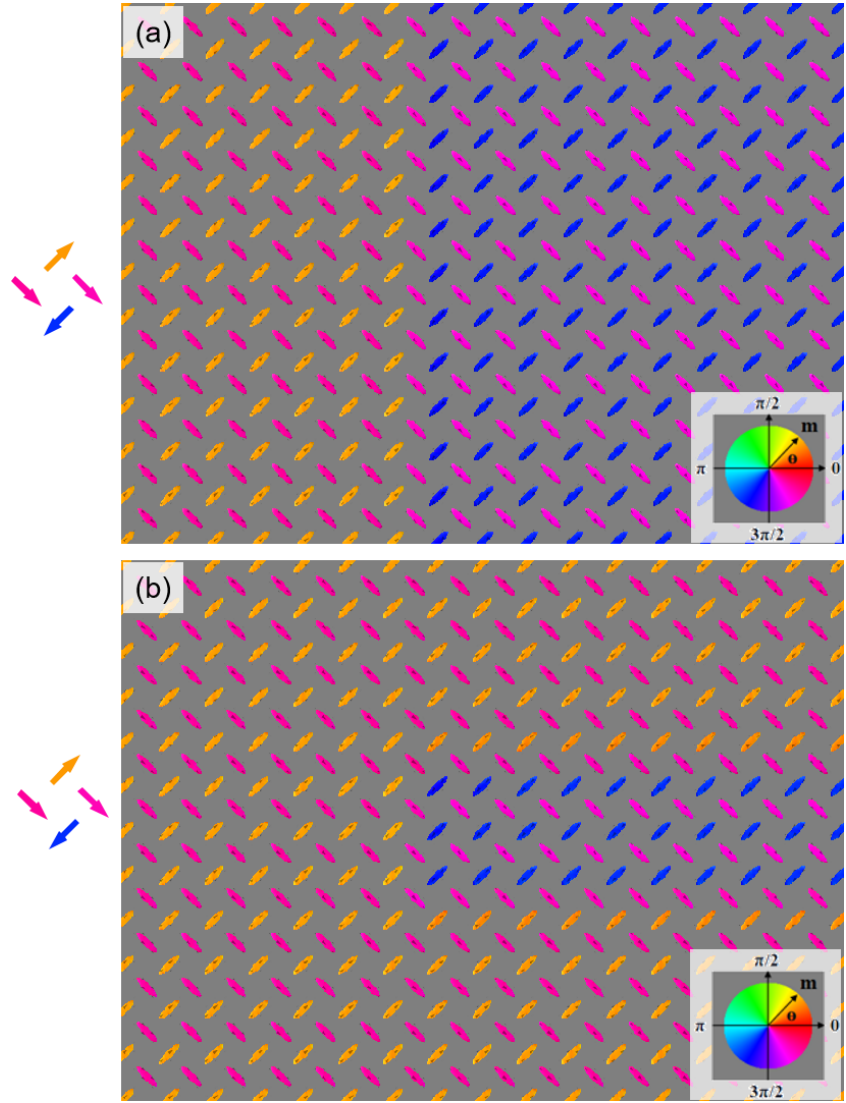

**Fig. S5 | Magnetic configurations of the top pinwheel ASI in the composite states. (a)** and **(b)** correspond to Figs. 4(a) and 4(b), respectively. The color encoded moment directions of the nanomagnets are shown by the arrows in the left.

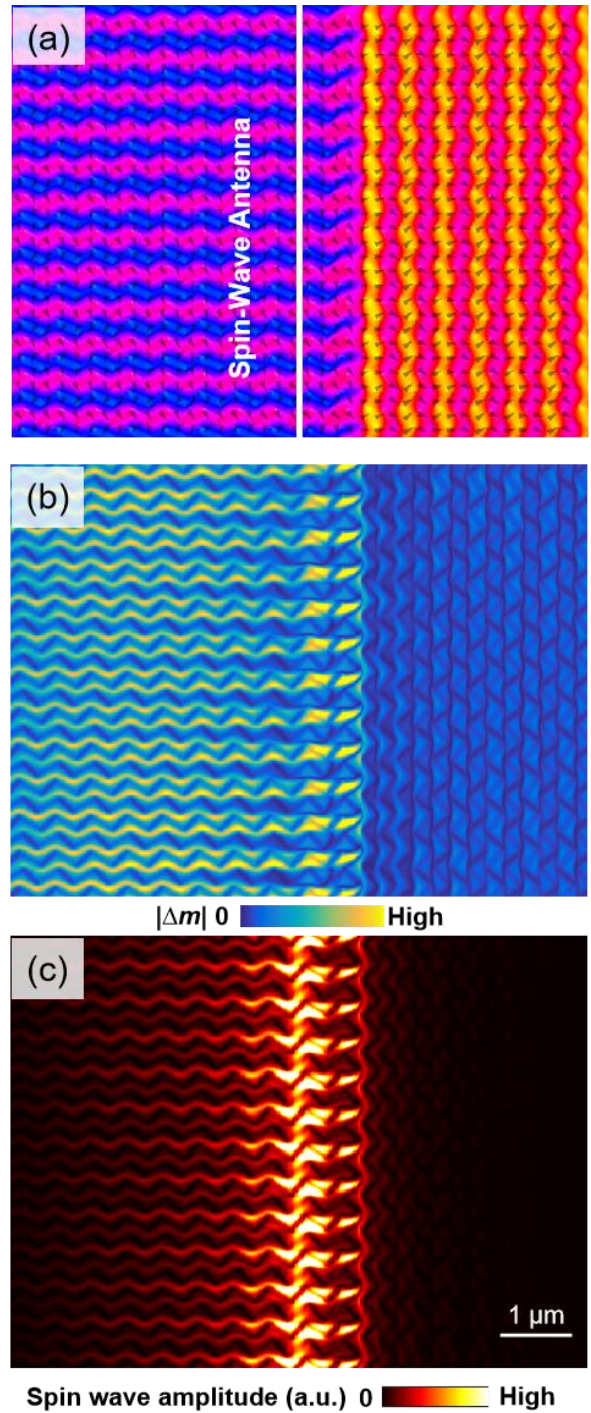

**Fig. S6 | Reversible one-way spin wave propagation. (a)** Composite magnetization state with a vertical domain state on the right of a horizontal domain state, which is opposite to that in Fig. 4(c). **(b)** and **(c)** Associated spatially resolved modes at $H_1$ = 3.71 GHz and spin wave amplitude map of propagating spin waves. The spin waves are excited from the white center line shown in (a). Scale bar, 1 μm.

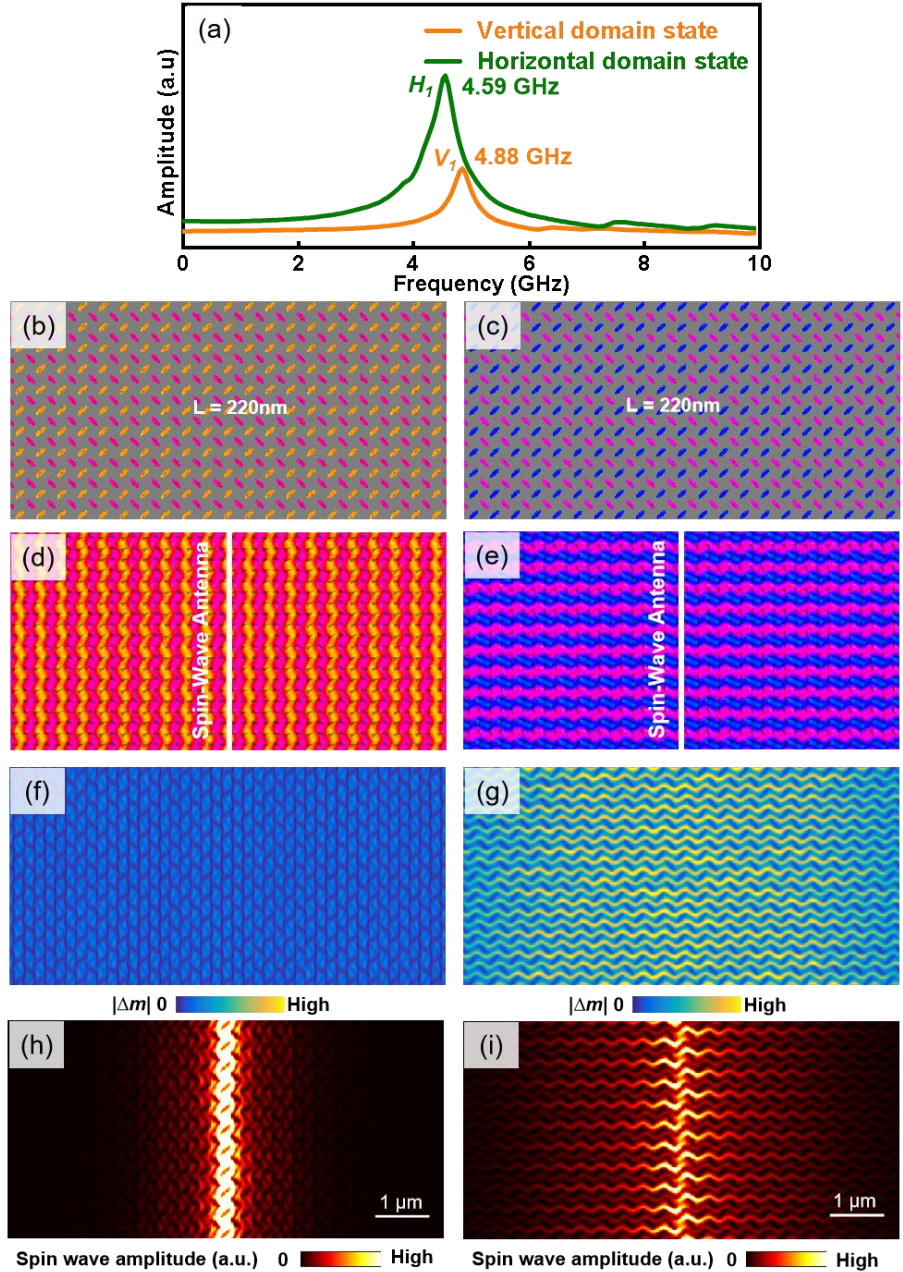

**Fig. S7 | Switchable spin wave propagation in the top pinwheel ASI comprised of elliptical nanobar magnets with dimensions of length $L$ = 220 nm, width $W$ = 80 nm and thickness $T$ = 20 nm.** **(a)** The spectra for vertical and horizontal domain states excited by a field pulse along the $x$ (horizontal) direction. **(b)** and **(c)** Magnetic configurations of the top pinwheel ASI of vertical and horizontal domain states. **(d)** and **(e)** Micromagnetic configurations of the underlayer film for vertical and horizontal domain states. **(f)** and **(g)** Spin wave profile of $H_1$ = 4.59 GHz corresponding to (d) and (e). **(h)** and **(i)** Amplitude maps of propagating spin waves at $f$ = 4.59 GHz, corresponding to the states in (d) and (e), respectively. The spin waves are excited from the white center line shown in (d) and (e).

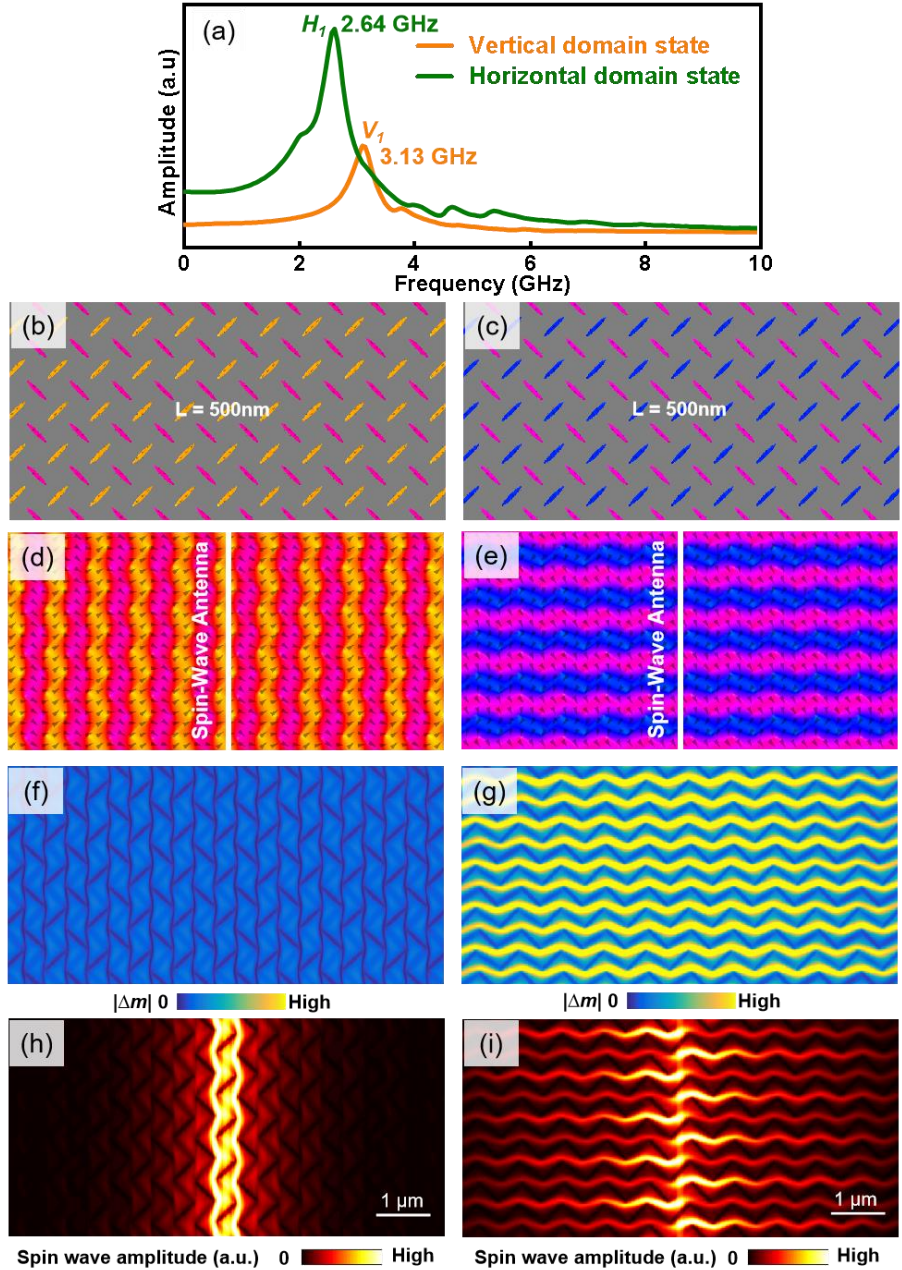

**Fig. S8 | Switchable spin wave propagation in the top pinwheel ASI comprised of elliptical nanobar magnets with dimensions of length $L$ = 500 nm, width $W$ = 80 nm and thickness $T$ = 20 nm.** (a) The spectra for vertical and horizontal domain states excited by a field pulse along the $x$ (horizontal) direction. (b) and (c) Magnetic configurations of the top pinwheel ASI for vertical and horizontal domain states. (d) and (e) Micromagnetic configurations of the underlayer of vertical and horizontal domain states. (f) and (g) Spin wave mode profile at $H_1$ = 2.64 GHz corresponding to (d) and (e). (h) and (i) Amplitude maps of propagating spin waves at $f$ = 2.64 GHz, corresponding to the states in (d) and (e), respectively. The spin waves are excited from the white center line shown in (d) and (e).

**The description and sequence of videos:**

Video1. Spatially resolved modes with varying frequencies for the vertical domain state in Fig. 2(c).

Video2. Spatially resolved modes with varying frequencies for the horizontal domain state in Fig. 2(c).

Video3. Spin wave propagation in the vertical domain state in Fig. 3(a).

Video4. Spin wave propagation in the horizontal domain state in Fig. 3(c).

Video5. Spin wave propagation in the composite state in Fig. 4(a).

Video6. Spin wave propagation in the composite state in Fig. S6(a).

Video7. Spin wave propagation in the composite state in Fig. 4(b).